\nonstopmode
\documentclass[prl,aps,superscriptaddress,twocolumn]{revtex4}
\usepackage{amsmath}
\usepackage{amssymb}
\usepackage{comment}
\usepackage{dsfont}
\usepackage{mathrsfs}
\usepackage{mathtools}
\usepackage{graphicx}
\usepackage{color}
\usepackage{xcolor}
\newcommand{\I}{\imath}

\newcommand{\ket}[1]{|#1\rangle}

\renewcommand{\Im}{\text{Im}~}

\usepackage{letltxmacro}
\makeatletter

\begin{document}
\title{ 
Non-equilibrium Floquet steady states of time-periodic driven Luttinger liquids}
\author{Serena Fazzini}
\affiliation{Physics Department and Research Center OPTIMAS, Technische Universit\"at Kaiserslautern, 67663 Kaiserslautern, Germany}
\author{Piotr  Chudzinski}
\affiliation{School of Mathematics and Physics, Queen's University Belfast, BT7 1NN Belfast, UK}
\affiliation{Institute of Fundamental Technological Research, Polish Academy of
Sciences, 02-106 Warsaw, Poland}
\author{Christoph Dauer}
\affiliation{Physics Department and Research Center OPTIMAS, Technische Universit\"at Kaiserslautern, 67663 Kaiserslautern, Germany}
\author{Imke Schneider}
\affiliation{Physics Department and Research Center OPTIMAS, Technische Universit\"at Kaiserslautern, 67663 Kaiserslautern, Germany}
\affiliation{Institute of
Physics, Universit\"at Augsburg, 86135 Augsburg, Germany}
\author{Sebastian Eggert}
\affiliation{Physics Department and Research Center OPTIMAS, Technische Universit\"at Kaiserslautern, 67663 Kaiserslautern, Germany}
\begin{abstract}
Time-periodic driving facilitates a wealth of novel quantum states and quantum engineering.  The interplay of Floquet states and strong interactions is particularly intriguing,
which we study using  
time-periodic fields in a one-dimensional quantum gas, modeled by 
a Luttinger liquid with periodically changing interactions.
By developing a time-periodic operator algebra, 
we are able to solve and analyze the complete set of non-equilibrium steady states 
in terms of a Floquet-Bogoliubov ansatz and known analytic functions. 
Complex valued Floquet eigenenergies occur 
when integer 
multiples of the driving frequency approximately match twice the dispersion energy, 
which correspond to resonant states. 
In experimental systems of Lieb-Liniger bosons we
predict a change from powerlaw correlations to dominant collective
density wave excitations at the corresponding wave numbers as
the frequency is lowered below a characteristic cutoff.
\end{abstract}

\maketitle
{\it Introduction.}
\label{sec:intro}
Controlled time-periodic driving of quantum systems has recently pushed the development
of fascinating quantum phenomena such as topological phases \cite{topo1,topo2},
many body localization \cite{mbl}, 
cavity opto\-mechanics \cite{Baumann_2010, Mottl_2012, Landig_2015, Klinder_2015, Landig_2016, Deng_2014, Brennecke_2008, Kulkarni_2013, Piazza_2015},
Floquet time crystals \cite{timextal1,timextal2},
artificial gauge fields \cite{Lin_2009,Aidelsburger_2011, Lin_2011, Struck_2013, Hauke_2012,  Struck_2012}, 
transmission resonances \cite{thuberg2016,Reyes_2017,thuberg2017}, 
dynamic localization \cite{DinLoc1,DinLoc2,DinLoc3,DinLoc4,DinLoc5,DinLoc6},
pairing \cite{Rapp_2012,wang2020},
driven Bose-Einstein condensates \cite{Bretin_2004, Schweikhard_2004, Ramos_2008, 
Pollack_2010, Wang_2014,Greschner_2014,Meinert_2016,Arimondo_2012},
 and anyons \cite{Keilmann_2011, Greschner_2015, Tang_2015, Str_ter_2016, Lange_2016}. 
However, when complications from strong correlations and non-equilibrium physics
become intertwined, understanding the dynamics becomes very difficult. 
Theoretical progress has been made in the high frequency 
limit \cite{Eckardt2005,Eckardt,Eckardt_2017}, 
which is useful for Floquet engineering.  On the other hand, it is extremely rare 
to obtain full solutions
of time-periodically driven many-body
systems, which could give much needed 
insight in Floquet-induced strong correlations near resonances.

In this Letter we now provide the many-body eigenstate solution and report
resonance phenomena in one-dimensional (1D) interacting quantum systems
with time-periodically driven parameters.  Our analysis 
applies to time-periodic driving of generic 
Tomonaga-Luttinger liquids (TLL), which describe a large class of 
effectively 1D many-body systems \cite{Giamarchi} and can also be 
realized using ultra-cold gases \cite{Kinoshita_2004,Paredes2004,Vogler2013,decouple}. 
The time-evolution of an initially prepared state in a TLL has been 
calculated before \cite{Pollmann,Citro,Chudzinski,Pielawa,Graf_2010,Heyl,Kagan,ChudzinskiXXZ}, but much less is known about the nature 
of possible {\it non-equilibrium steady states} under periodic driving.
It is therefore desirable to obtain the full eigenbasis of the Floquet eigenvalue problem, 
which gives systematic information about all stable steady states and their
corresponding dominant correlations.  
We now obtain the explicit steady state solutions 
of the time-dependent Schr\"odinger equation of a quantum many-body system 
in terms of a time-periodic operator algebra, 
which not only allow a full analysis using closed analytic functions, but also show regions 
of instabilities in frequency and momentum space.  We therefore predict 
large-amplitude density waves at the characteristic wave vectors in trapped 
ultracold boson systems.

\begin{figure}
\includegraphics[width=.89\columnwidth]{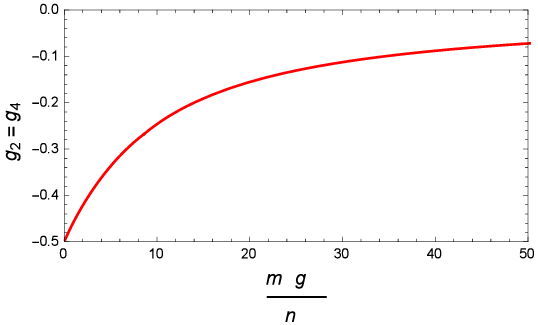}
\caption{The coupling constants $g_2=g_4 = (1/K^2 -1)/2$
for the Lieb-Liniger model as a function of  $mg/n$ \cite{cazalilla,Cazalilla_2004,Rist}, 
which can be determined for any value of $a_0$ and $a_\perp$.}
\label{K}
\end{figure}

{\it Model.}
\label{sec:model}
We will 
develop a Floquet-Bogoliubov ansatz 
for general driven TLL models. To make 
concrete predictions for decoupled 1D tubes of interacting bosons in 
optical lattices \cite{Kinoshita_2004,Paredes2004,Vogler2013,decouple} we choose the 
Lieb-Liniger Hamiltonian \cite{liebliniger} as a starting point
\begin{equation}
H_{0}=-\frac{1}{2m}\sum_{i=1}^N\frac{\partial^2}{\partial x_i^2}+g\sum_{i<j=1}^N\delta(x_i-x_j)
\label{lieb}
\end{equation}
where $\hbar=1$ and $g = \frac{2 a_0}{m a_\perp (a_\perp - 1.03 a_0)}$ is the  1D onsite interaction strength, 
which is tunable via the 3D scattering length
$a_0$
and the perpendicular confinement length $a_\perp$ \cite{confined,cazalilla,Cazalilla_2004,Rist}.  
The static system is integrable and correlations are known to be well described by a
TLL model in the long-wavelength limit $q<q_c$ \cite{Giamarchi,cazalilla,Cazalilla_2004} 
\begin{eqnarray} 
H_{TLL}&=& \frac{v}{2\pi}\int dx \left[ K\left(\partial_x\theta\right)^2+ K^{-1}\left(\partial_x\phi\right)^2\right] = \sum_{q>0} H_q \nonumber \\
H_q&=&  
v_F q \left[ (1+g_4)2 J_{0,q} + g_2 (J_{+,q} + J_{-,q})\right]
\label{TLL} 
\end{eqnarray}
where $2 J_{0,q}\! =\! b^\dagger_{L,q} b^{\phantom{\dagger}}_{L,q}\! +\! b^{\phantom{\dagger}}_{R,q} b^\dagger_{R,q}$ and $J_{+,q}\!=\! J_{-,q}^\dagger \!=\! b^\dagger_{L,q} b^{{\dagger}}_{R,q}$ 
are SU(1,1) generators \cite{ui,su11,su12} in terms of 
bosonic operators $b_{L/R, q}^\dagger$, which 
create left- and right-moving density
waves at wave-vector $q$ \cite{Giamarchi}. 
For the Lieb-Lininger model 
the Luttinger parameter $K$ is known exactly
\cite{cazalilla,Cazalilla_2004,Rist}.
 It depends only on the ratio $m g/n$,
 where $n$ is the 1D particle density.  
The cutoff $q_c$  above which the TLL description fails has also been determined \cite{Cazalilla_2004}.
The scattering amplitudes $g_2$ and $g_4$ 
are rescaled from the traditional ''g-ology'' scheme \cite{solyom} and 
related to $v$ and $K$ 
by \cite{Giamarchi}
\begin{eqnarray}
\label{vK}
 vK= 
v_F \left(1 + g_4 - g_2\right) , \ \ \ \ 
\frac{v}{K} = v_F\left(1+ {g_4} + {g_2}\right),
\end{eqnarray}
where $v_F = {\pi n}/{m}$. Values of \mbox{$g_2=g_4 = (1/K^2 -1)/2$} for the 
Lieb-Liniger model are shown in Fig.~\ref{K}.

We now turn to systems with time-periodically changing control parameters
$a_0$ and $a_\perp$, which will result in  
time-periodic couplings  $g$, $g_2$, and $g_4$, all of which can be determined exactly.  
Any desired time-periodic couplings
can be created by suitable fields given by the inverted relations, including 
a pure sinusoidal behavior \cite{note1}
\begin{eqnarray}
\label{coupl}
2 g_2(t) & = & 2 g_4(t) =   \bar \rho + \rho \cos \omega t
\end{eqnarray}
with constant parameters $\bar \rho$ and $\rho$.
We will later consider more general behavior.
For the Lieb-Liniger model it is known that $v K = v_F$ and $K>1$ \cite{cazalilla,Cazalilla_2004}, 
so that from Eq.~(\ref{vK}) $-1/2<g_2=g_4 <0$ as shown in Fig.~\ref{K}. 


{\it Floquet ansatz.}
\label{sec:solution}
We now seek to solve the time-dependent Schr\"odinger equation 
$\imath\partial_t|\Psi(t)\rangle=H_q(t)|\Psi(t)\rangle$
for each momentum $q$ separately, which remains a good quantum number and can be
omitted in the following.
According to Floquet theory \cite{Eckardt_2017,Eckardt,DinLoc4,Holthaus}  there exists a
complete set of quantum numbers $n$ for 
steady state solutions $|\Psi_n(t)\rangle=e^{-\I\epsilon_n t}|u_n(t)\rangle$.
Here $|u_n(t)\rangle=|u_n(t+T)\rangle $
with $T=2 \pi/\omega$ obey the Floquet equation
\begin{equation}
(H(t)-\I\partial_t)|u_n(t)\rangle=\epsilon_n|u_n(t)\rangle 
\end{equation}
where $\epsilon_n$ are the Floquet quasienergies.  We now wish to map this problem 
onto a {\it static} eigenvalue problem \cite{note2}
\begin{equation}
\tilde H |n\rangle =(Q H Q^{\dagger}-i Q\partial_t Q^{\dagger}) |n\rangle = \epsilon_n |n\rangle.
\label{diag}
\end{equation}
{\color{black}Floquet theory has been reviewed extensively 
\cite{Eckardt_2017,Eckardt,DinLoc4,Holthaus}, but the 
ansatz (\ref{diag}) goes beyond the usual 
time-evolution approach
since it makes the problem static, diagonalizes it in the original Hilbert space, 
and determines all steady states
for all times in one single unitary transformation $Q(t)$, which is an ambitious 
goal.  
The relation of $Q= \sum_n |n\rangle \langle u_n(t)|$ to Floquet concepts is
discussed in the Appendix:
While the time-evolution operator $W(t)$ is {\it not} the topic here, it can be
simply obtained $W(t)= Q^{\dagger}(t) e^{-i \tilde H t} Q(0)$. However, it is not possible
to construct $Q$ using $W$.  Likewise, the so-called Floquet Hamiltonian \cite{Eckardt_2017,Eckardt,DinLoc4,Holthaus} $H_F= Q^\dagger(0)\tilde H Q(0)$ can be found using $Q$.  
We now proceed to find an explicit expression for $Q(t)$ for the model in Eq.~(\ref{TLL}).

{\it Floquet Bogoliubov solution.} 
The goal is to find
a static eigenbasis in the rotating frame, which can be achieved if $\tilde H$ becomes
diagonal and time-independent.}
The interacting model $H_q(t)$ in Eq.~(\ref{TLL}) is defined in left and right oscillator Hilbert spaces $\chi= L,R$,  so 
a static solution must be of the form
$\tilde{H} = \Delta \sum_\chi b_{\chi}^{\dagger}b_{\chi}$. 
The characteristic commutation relation 
$[\tilde{H},b_{L,R}]=-\Delta b_{L,R}$ transforms to 
\begin{eqnarray}
\label{eq:eig_Fl}
[(H(t)-\I\partial_t),{\beta}_{L,R}(t)]& = & -\Delta \beta_{L,R}(t)\ {\rm with} \\
\beta_{L,R}(t)=Q^{\dagger}(t)b_{L,R}Q(t) & = & 
\gamma_1(t)b_{L,R}+\gamma_2(t)b_{R,L}^{\dagger} \label{beta}
\end{eqnarray}
where we have used a general Floquet-Bogoliubov ansatz for $Q$ in Eq.~(\ref{beta}) 
with the canonical 
constraint $|\gamma_1(t)|^2-|\gamma_2(t)|^2=1$. The 
defining relation in 
Eq.~(\ref{eq:eig_Fl}) provides 
differential equations for
the time-periodic
coefficients $\gamma_{1,2}$ 
\begin{equation}
\I \dot \gamma_{1,2} = (\Delta \mp\lambda_1)\gamma_{1,2} \pm \lambda_2 \gamma_{2,1} 
\label{eq:eig_vect}
\end{equation}
where $\lambda_1 = v_Fq(1+g_4)$ and $\lambda_2 = v_Fqg_2$.
The relation (\ref{eq:eig_vect}) applies to general TLL, but 
for the Lieb-Liniger model it simplifies since 
$\lambda_1-\lambda_2=q v_F$ is constant due to 
Galilean invariance. Using $f_\pm(t)=e^{\I\Delta t}(\gamma_1(t) \pm \gamma_2(t))$
and 
Eq.~(\ref{coupl}) we obtain a Mathieu equation
\begin{equation} \label{mathieu}
\ddot f_-(t) + q^2 v_F^2 (1+\bar \rho + \rho \cos \omega t ) f_-(t) = 0
\end{equation}
and $f_+ = -\I \dot f_-/q  v_F$.
The solution can be expressed as
\begin{eqnarray}
\label{eq:sol_Mathieu}
&& {f}_-(t)  =  
c_1 {\cal C}\left(a,p,\tau\right)+
c_2 {\cal S}\left(a,p,\tau\right)  \\
\rm{where} & &  a= 4 \frac{q^2 v_F^2}{\omega^2}(1+\bar{\rho}), \ \ p = -2 \frac{q^2 v_F^2}{\omega^2} \rho, \ \ \tau = \frac{\omega t}{2}, \nonumber
\end{eqnarray}
and ${\cal C}(a,p,\tau),\  {\cal S}(a,p,\tau)$ are even and odd Mathieu functions 
normalized with 
${\cal C}(a,p,0)={\cal S}(a,p,\pi)=1$.
 \begin{figure}
 \centering
\includegraphics[width=.89\columnwidth]{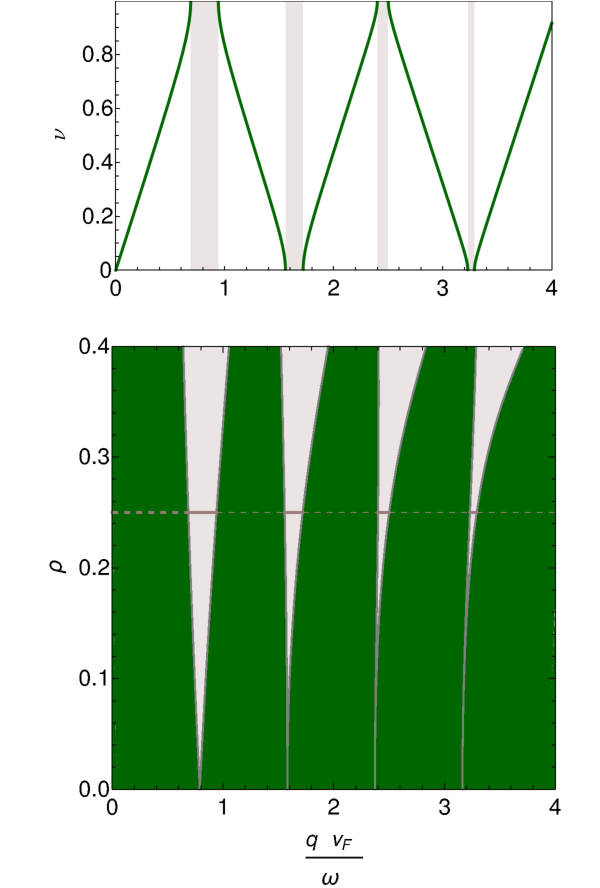}\\
 \caption{\label{fig:StabilityChart} {\it Top:} The value of $\nu=2\Delta/\omega$ 
as a function 
of $q v_F/\omega$ using $\bar \rho = -0.6$ and amplitude $\rho=0.25$.
Shaded regions indicate complex values of $\Delta$.
{\it Bottom:} Stability chart of the Mathieu equation with $\bar \rho=-0.6$. 
Grey areas are the instability regions
around the resonance points $q_\ell = \ell \omega/2\bar v$. }
 \end{figure}
{\color{black}
The coefficients $c_{1,2}$ are determined by the time-periodicity
of steady states $|u_n(t)\rangle$ and operators $\beta(t)$, which also fixes 
the quantization condition for $\Delta$:
We use Floquet's theorem to write the solution of Eq.~(\ref{mathieu})
$ {f}_{-}(t)=e^{\imath \nu \tau}P_{\nu}(\tau)$ 
with $P_{\nu}(\tau)=P_{\nu}(\tau+\pi)$ \cite{note3}. 
Since $\gamma_{1/2}$ 
are periodic, we find that the Mathieu characteristic exponent
is $\nu = 2 \Delta/\omega$, which must be real for 
stable steady states, just like for
the Mathieu stability chart \cite{StabilityChart} 
of Paul traps \cite{paul}.
From the normalization above follows
$\cos(\pi\nu)={\cal C}[a,p,\pi]$, which gives   (see Appendix) 
\begin{equation}
\Delta = \arccos[{\cal C}(a,p,\pi)]/T, \ \ \ \ \ 
c_2 = \imath c_1 \sin T \Delta,
\label{quantization}
\end{equation}
and $c_1$ is fixed by $|\gamma_1|^2-|\gamma_2|^2 = 1$.}
Last but not least, we can use the solutions of $\gamma_{1,2}$ to uniquely define three 
real time periodic functions $\theta, \phi, r$, which parametrize 
an explicit expression of $Q(t)$ in Eq.~(\ref{beta})
in terms of the SU(1,1) generators
$J_0$, $J_-$, and $J_+$ in Eq.~(\ref{TLL}) \cite{ui,su11,su12}
\begin{eqnarray} \label{Q}
Q(t) & =&  e^{i\theta J_0}e^{r(J_+-J_-)}e^{-i\phi J_0} \ {\rm with}\\
\gamma_1 & = & e^{i (\theta-\phi)/2}\cosh r \label{gamma11}, \ \ \ 
\gamma_2  =   e^{i(\theta+\phi)/2}\sinh r \label{gamma22}.
\end{eqnarray}
In the Appendix it is shown that $Q(t)$ in Eq.~(\ref{Q})
gives 
$\tilde H \!=\! \Delta ( b^\dagger_L b_L^{\phantom{\dagger}}\!\! +\! b_R^{\phantom{\dagger}} b_R^\dagger)$ 
and the form of the transformed
ground state $|u_0(t)\rangle \!\!=\!\! Q^{\dagger} |0\rangle$ is provided (see Appendix), which obeys $\beta_{L,R}(t) |u_0(t)\rangle = 0 \ \forall t$.  Therefore, from 
Eqs.~(\ref{eq:eig_Fl}) and (\ref{beta})
all Floquet modes $|u_n(t)\rangle$ with $\epsilon_n= (n_L+n_R+1) \Delta$
are found 
by application of $(\beta_{L}^\dagger(t))^{n_{L}}(\beta_{R}^\dagger(t))^{n_{R}}$ on $|u_0(t)\rangle$.


{\it Instability regions.} 
{\color{black}  Before calculating physical observables, we need to analyse 
the stability of the differential equations, 
which may not always have a solution due to the periodicity constraint.}
In Fig.~\ref{fig:StabilityChart}(top) we plot the value of $\nu=2\Delta/\omega$ 
as a function of 
rescaled momenta $q v_F/\omega$ using $\bar \rho = -0.6$ and amplitude $\rho=0.25$.  
We observe that for certain regions of momenta there are no real 
solutions for $\Delta$.  These ``instability regions'' 
will have
interesting physical
implications as discussed below. 
The stable regions
are shown as a function of amplitude $\rho$
in Fig.~\ref{fig:StabilityChart}(bottom) for $\bar \rho = -0.6$.
For small $\rho$ the
instability regions are equally spaced at integer values $\ell \in \mathbb N$
corresponding to $a = \ell^2$ or  $\ell = 2 \sqrt{1+\bar \rho} q v_F/\omega $.  Defining an average velocity $\bar v = v_F/\bar K = v_F \sqrt{1+\bar \rho}$ the instability regions therefore correspond to 
integer multiples of frequency which match twice the 
interacting dispersion relation $\ell  \omega = 2 \bar v q_\ell$, so the physical 
cause can be traced to resonant excitations on the linear branches
of left movers from $-\bar v q_\ell$ to right movers at $ \bar v q_\ell$ and vice versa.

\begin{figure}
\includegraphics[width=.79\columnwidth]{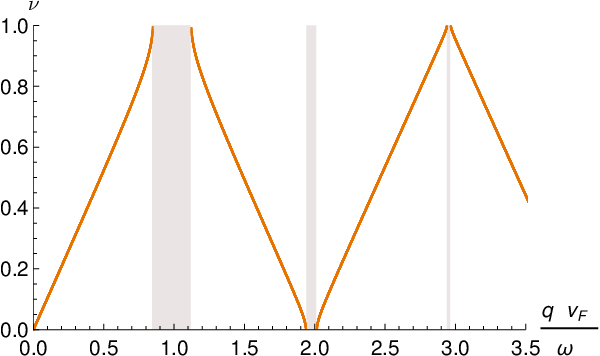}
\begin{picture}(0,0)
\end{picture}
\caption{Characteristic exponent $\nu$ for 
$g_4=-0.4$, while only $g_2(t)$
is driven with $\bar \rho=-0.6$ and $\rho = 0.25$ in Eq.~(\ref{coupl})}
\label{fig:num}
\end{figure}
{\color{black} As shown in Fig.~\ref{fig:num}, the region of instabilities also 
occur for more general TLL models where the restriction $g_2(t)=g_4(t)$
in Eq.~(\ref{coupl}) is lifted \cite{ChudzinskiXXZ}
and/or contain higher harmonics.  
A general analytic solution remains elusive, 
but the corresponding differential equation
(\ref{eq:eig_vect}) is still valid, which we have solved numerically by 
Fourier decomposition for several 
parameters.  
Instability regions are always 
expected since the problem is analogous to forbidden energy regions in a 
band structure of a periodic potential \cite{Holthaus}, which is of course generic.
In Fig.~\ref{fig:num} we show the behavior of $\nu$
as a function of $q v_F/\omega$ for the case that only the $g_2$ scattering process 
is periodically modulated in time with
$\bar \rho=-0.6$ and ${\rho}=0.25$ in Eq.~(\ref{coupl}) while $g_4 =-0.4$. While quantitative changes compared to 
Fig.~\ref{fig:StabilityChart}(top) can be identified, the regions of instabilities are again
found at resonant wave vectors.
In Figs.~\ref{fig:StabilityChart}(top) and \ref{fig:num}
we see that $\nu\to 1$ near the unstable regions
and the ratio $c_2/c_1$ in Eq.~(\ref{quantization}) becomes singular.}

\begin{figure}[t]
\includegraphics[angle=270,width=.99\columnwidth]{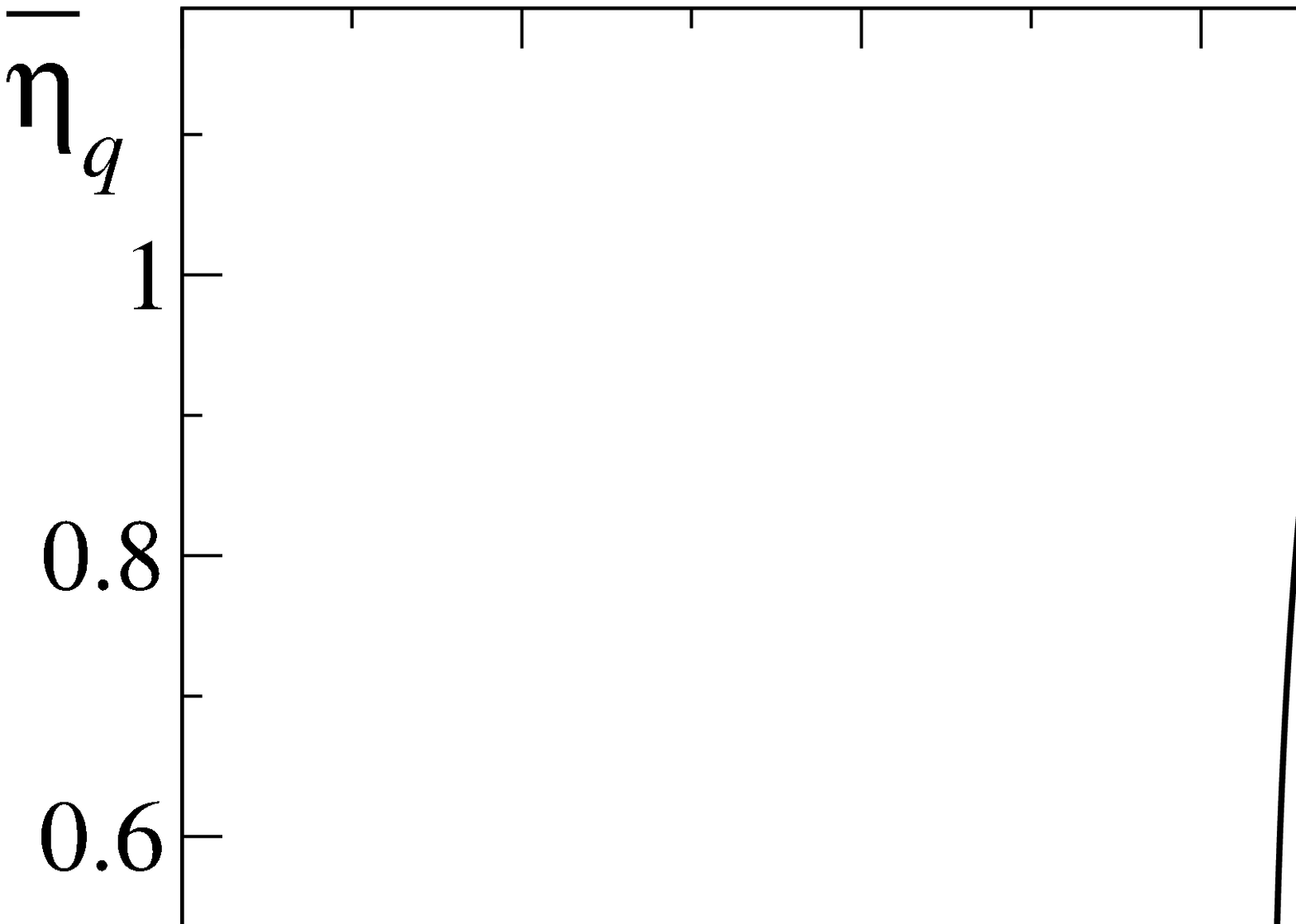}
\caption{\label{Fig:Occ}Time average of $\eta_q(t)$, plotted as a function of $q v_F/\omega$ for $\bar \rho=-0.6$ using different amplitudes ${\rho}$.
For $q\to 0$ the static limit 
${|\bar\gamma_{2}|^2}= (1/\bar K +{\bar K}-2)/4$ is recovered (red dot).
Inset: With finite life-time $\tau_0=10^4/v_Fq$ the divergent regions are turned into overwhelmingly 
large maxima.}
\end{figure}

{\color{black} 
\label{sec:results}
To understand the physical significance of the instability regions, 
it is essential to consider damping.   
Intrinsic damping is always 
present in the TLL description due to 
higher order boson-boson interaction terms \cite{Giamarchi}, which
lead to a finite quasiparticle life-time. 
A corresponding broadening of spectral peaks is seen numerically
for finite energies and in finite systems \cite{schneider,bohrdt}.
The size of damping is not universal since it depends on microscopic 
details including the system size, but it can be assumed to be smaller than
all other energy scales.
In Ref.~\cite{StabilityChart} it was shown that solutions of
{\it damped Mathieu equations} become always stable for amplitudes below a given threshold.
We also find that a finite life-time $\tau_0$ in form of an imaginary energy correction
$\Im \lambda_1 = -1/\tau_0$ leads to convergence of instabilities as discussed below.

{\it Results.}
We are now in the position to calculate physical observables.
The main effect of the time periodic driving is the 
excitation of density waves in the steady state. 
The number of density
excitations $b_{\chi q}^\dagger b_{\chi q}^{\phantom{\dagger}}$ ($\chi=L$ or $R$) 
in the transformed ground state 
$|u_0(t)\rangle$
is given by
\begin{equation}
\eta_{q}\! = \!\langle u_0(t)| b_{\chi q}^\dagger b_{\chi q}^{\phantom{\dagger}} | u_0(t) \rangle \! = \!
\langle 0|\beta^{\dag}_{\chi q}\beta_{\chi q}^{\phantom{\dagger}}|0 \rangle \!=\! |\gamma_2(t)|^2.
\end{equation} 
In Fig.~\ref{Fig:Occ} we plot the time average $\bar \eta_q$.  For small $q$
we find that $\bar \eta_q$ approaches the static limit, but a strong divergence
is observed as the instability region around $q_\ell$ is approached.  
In the inset of Fig.~\ref{Fig:Occ} we exemplarily show that a finite 
life-time $\tau_0=10^4/v_Fq$
turns the divergences of $\bar \eta_q$ into large maxima around $q_\ell$.
The height of the maxima can be tuned by the 
product $\rho\tau_0$. 

A universal physical picture emerges analogous to a resonance catastrophe:
A finite life-time has little effect away from resonance, but  
the resonance response is overwhelmingly large 
and proportional to $\tau_0$.  
If $q_\ell=\ell \omega/2 \bar v<q_c$ is in the TLL regime,
such maxima will therefore dominate the correlations.   
We find that $q_c \sim \bar v m/2$ is a good estimate for the cutoff. 

It is well known how TLL correlation are calculated \cite{Giamarchi}, which is reviewed
in the Appendix for the example of density-density correlations. 
An overwhelming maximum of $\bar \eta_q$ will 
dominate the correlations and lead to 
long-range density order (see Appendix)
\begin{equation}
\langle u_0| n(x) n(y) | u_0\rangle \propto \cos q_\ell (x-y). \label{density}
\end{equation}
For large driving amplitudes $\rho$ the
magnitude of the induced density waves can become larger than the background density, which 
may lead to fragementation into irregular density grains.}

{\it Discussion.}  
The three energy scales $\omega$, $\bar v q_c$, and  $v_Fq$ determine the behavior
of the system, which 
undergoes three different regimes as the frequency is changed:\\ {\bf 1.) High frequencies:}
For $\omega \agt \bar v q_c$ the instability regions are outside the TLL regime, so 
the physical relevant region is free of resonances. 
The transformation $Q$ results in 
a systematic change of $\bar \eta_q$ 
shown in Fig.~\ref{Fig:Occ}, which approaches the static limit as $q\to 0$.  The famous  
power-law correlations \cite{Giamarchi} are corrected for intermediate distances, but 
the asymptotic static limit is recovered.\\
{\bf 2.)  Intermediate frequencies:}  As the frequency is lowered, the resonant
wave-numbers $q_\ell = \ell \omega/2 \bar v$ drop below the cutoff $q_c$ into the 
TLL regime.
The number of density waves $\bar \eta_{q_\ell}$
becomes very large, dominating the correlations in Eq.~(\ref{density}).  Instead of powerlaw correlations, 
standing density 
waves at wave-numbers $q_\ell$ 
become stable throughout the system.\\
{\bf 3.) Very low frequencies:}  For $\omega\ll\bar v q_c$ extended regions 
of instability will 
lead to a large number of excitations and heating, destroying the correlations.

Using typical experimental parameters for a 1D $^{87}$Rb gas from Ref.~\cite{Vogler2013}
of $n = 6.2 \times 10^6/$m and $m g/n= 0.6$, we arrive at $\bar K \approx 4$ and a
cutoff of $\omega_c = \bar v q_c \approx 2\pi \times 1.4$kHz in the middle of the trap.
Driving the perpendicular confinement with a frequency of $\omega = 2\pi \times 500$Hz  
results in a resonance at $q_1 = \omega/2 \bar v = 444\times 10^3/$m.  
We therefore predict a standing density wave with wavelength $\lambda =2\pi/q_1$ in 
the $\mu$m range, which is
observable in real space with optical methods or 
an electron beam \cite{Vogler2013,decouple}.  

{\color{black} 
The confining trapping potential leads to lower local densities $n$ 
near edges \cite{Vogler2013,decouple}
and reduced velocities $v_F=\pi n/m$.  Everywhere $n$ agrees 
with the local density approximation (LDA)
of TLL correlations for the local trap potential
\cite{Vogler2013,decouple}.  The trapping potential is therefore turning into an advantage:
Instead of changing the frequency $\omega$, different regimes can be reached using
the changing density $n$.  As a function of  $n$ we know 
$\bar v = \sqrt{n g/m}$ \cite{cazalilla,Cazalilla_2004,Rist}, which in turn 
determines the resonant wave-vectors
$q_\ell = \ell \omega \sqrt{m/n g}/2$ and the cutoff $q_c = \sqrt{n g m}/2$.
Therefore, we move into the high frequency regime $q_\ell/q_c \propto \omega/ng$ as the 
density is lowered.
Note, that the density wavenumbers $q_\ell$ increase near edges 
in contrast to Fermionic 
Friedel density wavenumbers, which decrease with lower densities in a trap \cite{soeffing}.
In the proposed experiment, we therefore predict standing density 
waves at $\lambda \sim 14\mu$m
in the middle of the trap, which become shorter $\lambda \propto \sqrt{n}$
and weaker near the edge. 
It is an interesting open problem if significant
corrections would be observed when going beyond the present LDA analysis for a typical
trap size of 120 $\mu$m in Ref.~\cite{Vogler2013}.

Interesting many-body density excitations have been experimentally observed 
in driven 1D and 2D systems \cite{pattern2d,Li7}.
For 1D elongated bosonic $^7$Li gases
$\mu$m-size density grains emerge at $2\pi\times$80Hz driving,
which were identified as stable many-body effects \cite{Li7}.
Experimental images show grains that appear
smaller and weaker near edges 
which 
resemble features predicted above,
but in a random pattern \cite{Li7}. 
All correlations disappear for very low frequencies $\omega$.
A future grain size analysis as a function of $\omega$ and $n$ may
clarify if there is a relation to
TLL density waves in Eq.~(\ref{density}). }

{\it Conclusion.}  We have considered time-periodically driven interacting systems 
in the steady state, 
corresponding to generic TLL models in general or the Lieb-Liniger model in particular, 
which e.g.~applies to 1D confined atoms in ultra-cold gas experiments with tunable parameters.
As we have shown, this setup is one of the very rare cases where the combination of 
non-equilibrium steady states with many-body physics can be analyzed in great detail.
In particular, we have developed a Floquet-Bogoliubov approach by constructing 
time-periodic creation and annihilation operators, which solve the eigenvalue 
equation for the steady state 
by acting on the entire Floquet space.  We also identify regions in 
frequency-momentum space where damped resonant
behavior leads to a large number of density excitations.  The known static powerlaw correlations \cite{Giamarchi}
are recovered for large distances $\gg \bar v/\omega$, but for frequencies
below the cutoff $\bar v q_c$ characteristic density waves at integer-spaced resonant wave
numbers $q_\ell = \ell \omega/2 \bar v$ will become dominant.

We emphasize that the 
proposed Floquet-Bogoliubov algebra is completely general and can be used to solve
any time-periodically driven model with Bogoliubov-type interactions exactly.
The explicitly known transformation $Q$ maps all steady states onto a diagonal
static oscillator basis for all times, which 
paves the way for a
complete analysis of time-dependent effects in strongly interacting systems using 
a combination of powerful experimental, analytic, and numerical techniques.

\begin{acknowledgments}
We are thankful for support from Research Centers of
the Deutsche Forschungsgemeinschaft (DFG): 
Projects A4 and A5 in SFB/Transregio 185: ``OSCAR'' 
and Project A10 in SFB/Transregio 173: ``Spin+X''. 
\end{acknowledgments}

\section{appendix}
\label{app}

Here we give details on the Floquet Bogoliubov
transformation, its relation to Floquet theory, the explicit form of the transformed ground state state,
density-density correlations, and the application of Floquet's theorem to Mathieu functions.

\subsection{Relation of the time-dependent transformation to Floquet theory}
The goal is to find all possible steady state solutions 
$|u_n(t) \rangle=|u_n(t+T) \rangle$ under 
time-periodic driving
at each time $t$, which are defined by the Floquet eigenvalue equation
\begin{equation}
(H-i \partial_t)|u_n(t)\rangle = \epsilon_n | u_n(t)\rangle,
\label{floquet}
\end{equation}
where $\epsilon_n$ are real quasi energies.  It should be noted that 
it is not always possible to find steady state solutions, but if they exist they form
a complete basis in the original Hilbert space.
The underlying 
Floquet theory has been discussed in a number of review articles \cite{Eckardt_2017,Eckardt,DinLoc4,Holthaus},
where different approaches are presented:   By Fourier transforming into 
frequency space, the eigenvalue problem becomes static in an extended Hilbert space.
Different frequency components can be perturbatively decoupled using a Magnus
expansion, which is helpful in
defining a so-called Floquet Hamiltonian $H_F$.  The Floquet Hamiltonian is useful since it 
determines the quasi-energies and the stroboscopic time evolution. The eigenstates
of $H_F$ are the steady states $|u_n(0)\rangle$ {at one instant in time} only, so for
the full time evolution it is necessary to additionally know the micromotion operator 
$U(t)=\sum_n |u_n(t)\rangle \langle u_n(0)|$, which is in general more difficult.

Our novel approach is now to solve the Floquet eigenvalue problem in one single step
by mapping it to a static problem in the original Hilbert space
\begin{equation}
\tilde H | n\rangle = (Q H Q^{\dagger}-i Q\partial_t Q^{\dagger}) |n\rangle = 
\epsilon_n |n\rangle.
\label{diag2}
\end{equation}
If solutions to the original problem in Eq.~(\ref{floquet}) exist
the unitary transformation $Q$ can formally always be written as
\begin{equation}
Q(t) = \sum_n |n\rangle \langle u_n(t)|,
\label{trafo}
\end{equation}
which transforms the entire
basis of steady state solutions at each time into a diagonal static 
basis. 
This new transformation $Q$ therefore does three things at once: 
It maps the system to a static problem in the original Hilbert space, it diagonalizes the
eigenvalue problem, and it provides the time-dependent steady states for all times.
All this is done without using a Fourier transform into an extended Hilbert space. 
Needless to say, each of the above steps is normally highly non-trivial, so
finding such a transformation $Q$ into a diagonal rotating frame is very ambitious indeed.
Note, that $Q(t)=Q(t+T)$ is time periodic, but we need not assume that $Q(t)$ becomes the
identity at the initial time or any other time.

The operator $Q$ must therefore not be confused
with the time-evolution operator $W$ 
\begin{equation}
W(t) = \sum_n |u_n(t)\rangle \langle u_n(0)| e^{-i \epsilon_n t} = U(t) e^{-iH_Ft}, 
\end{equation}
which can be used 
to study the time-dependence of a given initial state.
In particular, knowing the time evolution cannot be used to construct $Q$, but 
the time evolution can always be expressed as
\begin{equation}
W(t) = Q^{\dagger}(t) e^{-i \tilde H t} Q(0).
\end{equation}
Moreover, the Floquet Hamiltonian can be obtained by 
$H_F = Q^{\dagger}(0)  \tilde H  Q(0)$, but again just 
knowing $H_F$ cannot be used to extract 
the steady states for all times unless $Q$ is known.  Finally, also the
micromotion operator $U(t) = Q^\dagger(t) Q(0)$ and all steady states
$|u_n(t)\rangle  = Q^\dagger(t) |n\rangle$ can be obtained with $Q$, 
so such a transformation truely contains a complete solution of the many-body driven system.

\subsection{Explicit form of the Floquet Bogoliubov transformation}
The model of interest can conveniently be expressed in terms of SU(1,1) 
generators
\begin{equation}
H(t) = \lambda_1 2 J_0  + \lambda_2 (J_+ + J_-), \label{H}
\end{equation}
where 
\begin{equation}
2 J_0 = b^\dagger_L b^{\phantom{\dagger}}_L + b^{\phantom{\dagger}}_{R} b^\dagger_{R}, \ \ \ J_+= J_-^\dagger = b^\dagger_L b^{{\dagger}}_{R},
\end{equation}
and $\lambda_1 = v_F q(1+g_4)$ and $\lambda_2= v_Fq g_2$ are the 
time-periodic coupling parameters.
For the static case it is known that the  transformation $U_1 = e^{r (J_+-J_-)}$ can be used
for diagonalization, using the following relations
for transformed operators 
$\tilde \Lambda = U_1\Lambda U_1^{\dagger}$
\cite{ui,su11,su12}
\begin{eqnarray}
 \tilde b_R & =& b_R \cosh  r - b_{L}^\dagger  \sinh r \label{rel1} \\
 \tilde b_{L} &=& b_{L} \cosh  r - b_{R}^\dagger  \sinh  r \\
 \tilde J_0& = &J_0 \cosh 2 r -  \frac{J_+ + J_-}{ 2} \sinh 2 r \\
 \tilde J_\pm& =& -J_0 \sinh 2 r + \frac{J_+ + J_-}{ 2} 
\cosh 2 r \pm\frac{J_+ - J_-}{ 2}\\
 \tilde J_+ &+& \tilde J_- =  -2 J_0 \sinh 2 r + (J_+ + J_-) \cosh 2 r 
\label{rel2}
\end{eqnarray}

For the time-dependent transformation, we need a more general ansatz parametrized
in terms of three real time-periodic parameters 
$\theta, \phi, r$
\begin{eqnarray}
\label{unitrafo}
Q(t) & =&  e^{i\theta J_0}e^{r(J_+-J_-)}e^{-i\phi J_0}\\
\label{unitrafoinverse}
Q^{\dagger} = Q^\dagger&  =&  e^{i\phi J_0}e^{-r(J_+-J_-)}e^{-i\theta J_0}.
\end{eqnarray}
Using relations Eqs.~(\ref{rel1})-(\ref{rel2}) together with gauge transformations, we
find that the
general time-dependent Bogoliubov transformation can be written as
\begin{eqnarray}
\beta_\chi = Q^{\dagger} b_\chi Q & = & \gamma_1 b_\chi + \gamma_2 b_{\bar \chi}^\dagger\label{bogo1}\\
Q b_\chi Q^{\dagger} & = & \gamma_1^* b_\chi - \gamma_2 b_{\bar \chi}^\dagger \label{bogo2}
\end{eqnarray}
with $\chi = L,R$ and
\begin{eqnarray}
\gamma_1 & = & e^{i (\theta-\phi)/2}\cosh r \label{gamma1}\\
\gamma_2 & = &  e^{i(\theta+\phi)/2}\sinh r \label{gamma2}
\end{eqnarray}
With this parametrization the transformed operators 
$\tilde \Lambda = Q \Lambda Q^{\dagger}$ can again be straightforwardly 
derived from Eqs.~(\ref{bogo1})-(\ref{gamma2}) 
\begin{eqnarray}
\label{rel3}
 \tilde J_0  & = & \cosh 2 r J_0 - \tfrac{1}{2} \sinh 2 r (e^{i\theta} J_+ + h.c.)\\
\tilde J_++\tilde  J_-  & = & -2 \cos \phi \sinh 2 r \ J_0 +\\
& & \! \! \! \left[(\cos \phi \cosh 2 r -i \sin \phi)   e^{i \theta} J_+ \! +\! h.c.\right]\\
-i Q \partial_t Q^{\dagger} & =&  (- \dot \theta  + \dot \phi \cosh 2 r) J_0 \\
& & + \left[(i \dot r -\tfrac{\dot\phi}{2}\sinh 2 r) e^{i \theta} J_+ + h.c.\right]
\label{rel4}
\end{eqnarray}
Note, that the three real parameters $\theta, \phi, r$ give a general one-to-one
parametrization of the complex functions 
$\gamma_1$ and $\gamma_2$ 
which obey $|\gamma_1|^2-|\gamma_2|^2=1$.  The functions $\gamma_1$ and $\gamma_2$
have been extensively discussed in the paper so the transformation $Q$ is 
already explicitly known, but what is left to show in the following 
is that the Hamiltonian in Eq.~(\ref{H}) indeed becomes static and
diagonal when using those functions.

The defining differential equation is given in Eq.~(9) of the paper 
in terms of $\gamma_1$ and $\gamma_2$
\begin{eqnarray}
i \dot \gamma_1 & =& (\Delta -\lambda_1)\gamma_1 + \lambda_2 \gamma_2 \\
i \dot \gamma_2 & =& (\Delta +\lambda_1)\gamma_2 - \lambda_2 \gamma_1 
\end{eqnarray}
where $\Delta$ is a real constant which is fixed by the constraint that 
both $\gamma_1$ and $\gamma_2$ are periodic as discussed in the paper.  
In terms of the parametrization $\theta, \phi, r$, the differential equations
become 
after multiplying by $\exp(-i \frac{\theta\pm\phi}{2})$ respectively
\begin{eqnarray}
i \dot r  \sinh r - \frac{\dot \theta-\dot\phi}{2}\cosh r& = &  (\Delta-\lambda_1) \cosh r + \lambda_2 e^{i\phi} \sinh r  \nonumber\\
  i \dot r \cosh r  - \frac{\dot \theta+\dot\phi}{2} \sinh r &  =&
(\Delta+\lambda_1) \sinh r - 
\lambda_2 
e^{-i\phi} 
\cosh r \nonumber
\end{eqnarray}
The imaginary parts of both equations give the same relation
\begin{equation}
\dot r =\lambda_2\sin \phi  \label{imag}
\end{equation}
The real parts give
\begin{eqnarray}
\!\!\!\!0  & =& (\Delta + \dot \theta/2 -\lambda_1-\dot \phi/2)\cosh r+\lambda_2\cos\phi \sinh r\label{10}\\
\!\!\!\!0  & =& (\Delta +\dot \theta/2 +\lambda_1+\dot \phi/2)\sinh r -\lambda_2\cos\phi \cosh r
\label{11}\end{eqnarray}
For later use we take (\ref{10})$\times\sinh r  - $(\ref{11})$\times\cosh r $, which gives
\begin{eqnarray}
0  & =& -(\lambda_1+\dot \phi/2)\sinh 2 r+\lambda_2\cos\phi \cosh 2 r \label{offdiag}
\end{eqnarray}
Likewise (\ref{11})$\times\sinh r  - $(\ref{10})$\times\cosh r $ gives
\begin{eqnarray}
\!\!\!\!\Delta  & =& -\dot\theta/2 +(\lambda_1+\dot \phi/2)\cosh 2 r-\lambda_2\cos\phi \sinh 2 r
\label{diagpart}
\end{eqnarray}

We now turn to identify the different parts in the transformed 
Hamiltonian 
\begin{equation}
\tilde H= QHQ^{\dagger} -i Q \partial_t Q^{\dagger}
\end{equation}
Collecting all the terms of $\tilde H$ from Eqs.~(\ref{rel3})-(\ref{rel4})
we find that 
the prefactor of the diagonal part $2 J_0$  
reads
\begin{equation}
(\lambda_1+ \tfrac{\dot \phi}{2}) \cosh 2 r - \lambda_2\cos\phi \sinh 2 r -\tfrac{\dot \theta}{2} 
\end{equation}
which is exactly $\Delta$ according to Eq.~(\ref{diagpart}) and therefore time-independent.
The prefactor of the off-diagonal part $e^{i \theta} J_+$ is given by 
\begin{equation}
-\lambda_1 \sinh 2 r +\lambda_2(\cos \phi \cosh 2 r -i \sin \phi) + i \dot r -\tfrac{\dot\phi}{2}\sinh 2 r.
\end{equation}
Using Eq.~(\ref{imag}) for the imaginary part and Eq.~(\ref{offdiag}) for the 
real part, we see that this expression is indeed zero, so that we have shown that
the model in Eq.~(\ref{H}) transforms to 
\begin{equation}
\tilde H= QHQ^{\dagger} -i Q \partial_t Q^{\dagger} = 2\Delta  J_0 = \Delta
( b^\dagger_L b_L^{\phantom{\dagger}}\! +\! b_R^{\phantom{\dagger}} b_R^\dagger)
\end{equation}
where the constant $\Delta$ is determined by the constraint of periodicity 
and Floquet's theorem as described in the text.

\subsection{The transformed ground state}
We give an explicit expression of the transformed ground state $\ket{u_0(t)}=Q^{\dagger}\ket{0}$ and show that it indeed satisfies the  condition
\begin{equation}
\label{eq:AVVacuumCondition}
\beta_{L,R}(t)\ket{u_0(t)}=0 \ \ \forall t.
\end{equation}
With Eq.~\eqref{unitrafoinverse} the calculation of $Q^{\dagger}\ket{0}$ is split into three steps, one for each operator exponential. As $\ket{0}$ is an eigenstate of $J_0$, the first step yields $e^{-i\theta J_0}\ket{0}=e^{-i \theta/2}\ket{0}$. Using the relation \cite{su11}
\begin{equation}
e^{-r(J_+-J_-)}=e^{-\tanh(r)J_+} e^{-2 \ln(\cosh(r))J_0}e^{\tanh(r)J_-}
\end{equation}
and $J_-\ket{0}=0$, we find as an intermediate result
\begin{equation}
Q^{\dagger}\ket{0}=e^{-i \theta/2}e^{i\phi J_0}e^{-\tanh(r)J_+}e^{-\ln(\cosh(r))}\ket{0},
\end{equation}
With the definition of $\gamma_{1}$ and $\gamma_{2}$ in Eqs.~(\ref{gamma1}) and (\ref{gamma2}) we further simplify $e^{-\ln(\cosh(r))}=1/|\gamma_1|$ and $\tanh(r)=|\gamma_2|/|\gamma_1|$. The action of the last part of the transformation is found to be 
\begin{equation}
e^{i \phi J_0}e^{-|\gamma_2|/|\gamma_1| ~J_+}\ket{0}=e^{i \phi/2}\sum_{n=0}^{\infty} (-|\gamma_2|/|\gamma_1|e^{i \phi})^n \ket{n}_{\rm L}\ket{n}_{\rm R}.
\end{equation}
With $e^{i \phi}|\gamma_2|/|\gamma_1|=\gamma_2/\gamma_1$ 
we finally find an explicit expression for the transformed ground state
\begin{equation}
\label{eq:AVTransformedVacuum}
\ket{u_0(t)}=\frac{1}{\gamma_1}e^{-\frac{\gamma_2}{\gamma_1} b_L^\dagger b_R^\dagger}\ket{0}.
\end{equation}
It is important to note that while the form of state \eqref{eq:AVTransformedVacuum} is similar to the results of a static Bogoliubov transformation \cite{su12} here all parameters are time-dependent.
Using the transformation $Q$ the state \eqref{eq:AVTransformedVacuum} solves the Floquet Eq.~(5) in the main article with $\epsilon_0=\Delta$.  Moreover, we can show explicitly 
that the transformed ground state $|u_0(t)\rangle$
obeys condition Eq.~\eqref{eq:AVVacuumCondition} by applying $\beta_{L}(t)=\gamma_1(t) b_{L}+\gamma_2(t)b^\dagger_{R}$ to Eq.~\eqref{eq:AVTransformedVacuum}, which reads
\begin{equation}
\label{eq:AVExpliciteVacuumCondition}
\begin{split}
\beta_{L}(t)\ket{u_0(t)}=\ \ \ \ \ \ \ \ \ \ \ \ \ \ \ \ \ \ \ \ \ \ \ \ \ \ \ \ \ \ \ \ \ \
 \ \ \ \ \ \ \ \ \ \ \ \ \ \ \ \ \ \  \\
\frac{1}{\gamma_1}\sum_{n=0}^{\infty}\left(-\frac{\gamma_2}{\gamma_1}\gamma_1+
\gamma_2\right)\left(-\frac{\gamma_2}{\gamma_1}\right)^n \sqrt{n+1} \ket{n}_L\ket{n+1}_R.
\end{split}
\end{equation}
As the first bracket in \eqref{eq:AVExpliciteVacuumCondition} vanishes trivially, the state \eqref{eq:AVTransformedVacuum} is indeed the ground state of the $\beta_{L}(t)$ operator obeying Eq.~\eqref{eq:AVVacuumCondition} and analogously also for $\beta_{R}(t)$. This is an important result, as $|u_0(t)\rangle$ serves as base case for generating the entire set of steady states $|u_n(t)\rangle$
by application of $(\beta_L^\dagger(t))^{n_L}(\beta_R^\dagger(t))^{n_R}$ using Eq.~(7) in the main article. 

\subsection{Correlation functions}
It is well known how to calculate correlation functions of physical operators in
terms of the diagonal boson model $\tilde H$ \cite{Giamarchi,cazalilla,Cazalilla_2004}.
Of particular interest for ultra-cold gases is the density-density correlation, which we
will consider here to exemplify the calculation.  The fluctuating density
is given in terms of the bosonic field
$n(x) = \partial_x \phi(x)/\pi$, which has the mode expansion \cite{Giamarchi,cazalilla,Cazalilla_2004}
\begin{equation}
\partial_x \phi = \sum_{q>0} \left[\sqrt{\frac{\pi q}{2 L}} e^{i q x}\left(b_{L,q}^\dagger + 
b^{\phantom{\dagger}}_{R,q}\right) + h.c.\right]
\end{equation}
For the density-density correlation function we find in the transformed ground state
$| u_0(t)\rangle = Q^\dagger(t) | 0\rangle$
\begin{equation}
\langle u_0| n(x) n(y) | u_0\rangle
= \sum_{q>0}\frac{q}{L\pi} |\gamma_1 + \gamma_2^*|^2 \cos q(x-y) \label{sum}
\end{equation}
where we have used Eq.~(\ref{bogo1}).  If the parameters $\gamma_{1,2}$ are constant we recover
the known asymptotic powerlaw behavior $\frac{1}{2 \pi^2}|\gamma_1 + \gamma_2^*|^2/|x-y|^2$ \cite{Giamarchi,cazalilla,Cazalilla_2004}.  However, if a resonance $q_\ell = \ell \omega/2 \bar v$ is part of the linear
TLL regime, the parameters $\gamma_{1,2}$ will become very large as 
discussed in the main article. 
Therefore, the  
sum in Eq.~(\ref{sum}) will be dominated by the corresponding 
instability region, leading to a long-range density order of the form  
\begin{equation}
\langle u_0| n(x) n(y) | u_0\rangle \propto \cos q_\ell (x-y).
\end{equation}

\subsection{Floquet solution in terms of Mathieu functions}
The solution of the Mathieu equation
\begin{equation}
\ddot{y}(\tau)+(a-2p\cos{2 \tau})y(\tau)=0
\end{equation}
is usually discussed in terms of even and odd solutions, known respectively as Mathieu cosine ${\cal C}$ and Mathieu sine ${\cal S}$ functions. A general solution can be therefore written as
\begin{equation}
\label{SM:Mat_Sol}
y(\tau)=c_1 {\cal C}(a,p,\tau)+c_2{\cal S}(a,p,\tau) \hspace{2pt},
\end{equation}
with $\tau=\omega t/2$.
Floquet's theorem states that the solutions of a time-periodic differential equation 
can always be written in the form
\begin{equation}
\label{SM:Fl_Sol}
y(\tau)=e^{\imath\nu\tau}P_{\nu}(\tau) 
\end{equation} 
with $P_{\nu}(\tau)=P_{\nu}(\tau\pm\pi)$. We want to use the quantum number $\nu$, which is commonly referred to as Mathieu characteristic exponent. Therefore, in this section we clarify the relation between the latter and the Mathieu functions. 
Comparing Eqs.~(\ref{SM:Mat_Sol}) and (\ref{SM:Fl_Sol}) and employing the periodicity of $P_{\nu}(\tau)$, we get the following relation 
\begin{equation}
\begin{split}
c_1{\cal C}(a,p,\tau)+c_2&{\cal S}(a,p,\tau)=\\
&e^{\mp\imath \nu\pi}\left(c_1{\cal C}(a,p,\tau\pm\pi)+c_2{\cal S}(a,p,\tau\pm\pi)\right)\hspace{2pt}.
\end{split}
\end{equation}
Evaluating this expression in $\tau=0$ and normalizing the Mathieu functions such that ${\cal C}(a,p,0)={\cal S}(a,p,\pi)=1$, we obtain 
\begin{equation}
c_1(e^{\pm\imath\pi\nu}-{\cal C}(a,p,\pi))=\pm c_2 {\cal S}(a,p,\pi) = \pm c_2 \hspace{2pt},
\end{equation}
from which we finally get
\begin{equation}
\begin{split}
&\cos{\pi\nu}={\cal C}(a,p,\pi)\hspace{2pt},\qquad\text{and}\\
& c_2=\imath c_1\sin{\pi\nu} \hspace{2pt}.
\end{split}
\end{equation}

\subsection{Static Bogoliubov transformation}
For the time-independent case, Hamiltonian (2) in the main article can be expressed as 
\begin{equation}
H_q=  v_F q\left[(1+g_4)\left(2 J_{0,q}-1\right)+ g_2 \left(J_{+,q}+J_{-,q}\right)\right] 
\end{equation}
where  $2J_{0,q}=b_{Rq}^{\dagger}b_{Rq}+b_{Lq}b_{Lq}^{\dagger}$, $J_{+,q}=b_{Rq}^{\dagger}b_{Lq}^{\dagger}$, $J_{-,q}=J_{+,q}^{\dagger}$. For the sake of simplicity, in the following we will drop the index $q$. We observe that $J_i (i=0,\pm)$ form a $su(1,1)$ algebra; therefore the Tomonaga-Luttinger Hamiltonian can be diagonalized by the Schrieffer-Wolff transformation obtained through the unitary operator
\begin{equation}
\label{eq:Schrieffer_Wolff}
U=e^{z}\qquad z=\vartheta\left(J_+-J_-\right)
\end{equation}
\begin{equation}
\begin{split}
 H & \longrightarrow \tilde{H}=UHU^{\dagger}\\
|\rm{GS}\rangle & \longrightarrow |\widetilde{\rm{GS}}\rangle=U|\rm{GS}\rangle \hspace{2pt}.
\end{split}
\end{equation}
The transformed Hamiltonian $\tilde{H}$ is diagonal in the old bosonic basis
\begin{equation}
\begin{split}
\tilde{H}&=\Delta\left(b_R^{\dagger}b_R+b_L^{\dagger}b_L\right)+\Delta- v_F q (1+g_4)\\
&=2\Delta J_0- v_F q (1+g_4) \hspace{10pt}
\end{split}
\end{equation} 
with $\Delta= v_F q\sqrt{(1+g_4)^2-g_2^2}$ if $\tanh{(2\vartheta)}=\tanh{(2\bar{\vartheta})}\equiv g_2/(1+g_4)$, implying $\cosh(2\bar{\vartheta})=v_Fq(1+g_4)/\Delta$, $\sinh(2\bar{\vartheta})=v_Fqg_2/\Delta$.\\
Notice that we have used a passive transformation, where the operators are rotated, while the states defined by the original creation and annihilation operators stay the same. 
  \subsection{Classical model}
  The Hamiltonian
  \begin{equation}
  H(t)=(A+B(t))\left(a^{\dagger}a+b^{\dagger}b\right)+C(t)\left(ab+a^{\dagger}b^{\dagger}\right)
  \end{equation}
  with real coefficients $A, B(t), C(t)$ can be mapped to the classical model
  \begin{equation}\label{eq:Ham_cl}
  H(t)=\frac{1}{2}(A+B(t))\left(x^2+y^2+p_x^2+p_y^2\right)+C(t)\left(xy-p_xp_y\right)
  \end{equation}
  with the substitution
  \begin{equation}
  a=\frac{1}{\sqrt{2}}\left(x+\imath p_x\right)\qquad b=\frac{1}{\sqrt{2}}\left(y+\imath p_y\right) \hspace{2pt}.
  \end{equation}
  The Hamilton's equations for Hamiltonian (\ref{eq:Ham_cl}) are
  \begin{eqnarray}
  &\dot{x}=\frac{\partial H}{\partial p_x}=\left(A+B(t)\right)p_x-C(t)p_y \label{eq:Hamilton_x}\\
  &\dot{y}=\frac{\partial H}{\partial p_y}=\left(A+B(t)\right)p_x-C(t)p_x \label{eq:Hamilton_y}\\
  &\dot{p_x}=-\frac{\partial H}{\partial x}=-\left(A+B(t)\right)x-C(t)y \label{eq:Hamilton_px}\\
  &\dot{p_y}=-\frac{\partial H}{\partial y}=-\left(A+B(t)\right)y-C(t)x \label{eq:Hamilton_py} \hspace{2pt}.
  \end{eqnarray}
  By summing Eqs.~(\ref{eq:Hamilton_x}) and (\ref{eq:Hamilton_y}), and Eqs.~(\ref{eq:Hamilton_px}) and (\ref{eq:Hamilton_py}), we get the system
  \begin{eqnarray}
  &\dot{v}=\left(A+B(t)-C(t)\right)p_v\\
  &\dot{p_v}=-\left(A+B(t)+C(t)\right)v
  \end{eqnarray}
  for $v=x+y$ and $p_v=p_x+p_y$, which yields
  \begin{equation}
  \ddot{v}-\frac{\dot{B}(t)-\dot{C}(t)}{A+B(t)-C(t)}\dot{v}+\left[\left(A+B(t)\right)^2-C^2(t)\right]v=0 \hspace{2pt}.
  \end{equation}
  In the special case $C(t)=B(t)=\bar{\rho}+\rho\cos(\omega t)$, one recovers the Mathieu equation
  \begin{equation}
  \ddot{v}+\left[A^2+2A\bar{\rho}+2A\rho\cos(\omega t)\right]v=0 \hspace{2pt}.
  \end{equation}
The classical equations of motions also give insight to the role of damping,
  which generally reduces the regions of instability of the Mathieu equation.
In fact, in \cite{StabilityChart} it is shown that the presence of  linear damping 
  pushes the instability zone upwards in Fig.~2, so that below a critical value of amplitude
 $\rho$
stable solution are always possible and no instabilities occur.  
In \cite{StabilityChart} it is also shown that nonlinear effects can generate subharmonic stable motions. These cubic terms correspond to band curvature in the original band structure, so
   in a real system the nonlinearity of the band further stabilizes the 
system.  While a large number of density waves is still expected to occur at the critical $q-$values, the sum over all momenta becomes well defined leading to the predicted 
density order of scenario 2) in the main paper.
 \subsection{Floquet theory}
 Analogously to the Bloch theorem, the Floquet theorem asserts that the Schr\"odinger equation for a time-periodic Hamiltonian admits steady-state solutions of the form 
 \begin{equation}
 |\Psi(t)\rangle=e^{-\I\epsilon t}|u(t)\rangle  \hspace{2pt},
 \end{equation}
 where the modes $|u(t)\rangle=|u(t+T)\rangle $ inherit the periodicity from the Hamiltonian, and the quantity $\epsilon$ is the so-called Floquet quasienergy. Indeed the steady-state Schr\"odinger equation can be recasted in the form of an eigenvalue equation for the quasienergy operator $\mathscr{H}=H(t)-\I\partial_t$ in the extended Hilbert space generated by the product of the state space of the quantum system and the space of square-integrable $T$-periodic functions:
 \begin{equation}
 \label{eq:eig}
 \mathscr{H}|u(t)\rangle=\epsilon|u(t)\rangle \hspace{2pt}.
 \end{equation}
 By expanding both the Hamiltonian and the Floquet mode in Fourier series 
 \begin{gather}
 H(t)=\sum_m e^{\imath m \omega t}H^{(m)} \hspace{2pt},\\
 |u(t)\rangle=\sum_m e^{\imath m \omega t}|u_m\rangle \hspace{2pt},
 \end{gather}
 Eq. (\ref{eq:eig}) yields
 \begin{equation}
 \label{eq:Fl_components}
 \left(H^{(0)}+m\omega\right)|u_m\rangle +H^{(1)}\left(|u_{m-1}\rangle+|u_{m+1}\rangle\right)=\epsilon|u_m\rangle \hspace{2pt},
 \end{equation}
 which turns out to be an eigenvalue equation for the infinite tridiagonal matrix
 \begin{equation}
 \label{eq:matr1}
 {\cal M}_{\cal F}=\left(\begin{array}{ccccc}
   \vdots & \vdots  & \vdots  & \vdots  & \\
 \hdots & H^{(0)}-\omega & H^{(1)}  & 0 &  \hdots \\
 \hdots & H^{(1)} & H^{(0)} & H^{(1)}&  \hdots\\
 \hdots  & 0 & H^{(1)} & H^{(0)}+\omega  & \hdots\\
  & \vdots  & \vdots  & \vdots  &
 \end{array}
 \right) \hspace{2pt}.
 \end{equation}


\begin{thebibliography}{80}
\bibitem[{Note1()}]{note1}
\bibinfo{note}{The applied fields are not necessarily purely sinosoidal,
  but $a_0(g_2(t))$ or $a_\perp(g_2(t))$ can be obtained exactly for a given
  situation and therefore also the required fields. For small amplitudes a
  linear expansion will generally also result in sinosoidal fields.}
\bibitem[{Note3()}]{note3}
\bibinfo{note}{Here, $\nu$ is also analogous to the quasimomentum in the
  well-known bandstructure of a cosine-potential \cite{Holthaus}}.
\end{thebibliography}
\end{document}